\documentclass[preprint,showpacs,preprintnumbers,amsmath,amssymb,eqsecnum]{revtex4}
\usepackage{graphicx}
\usepackage{dcolumn}
\usepackage{bm}
\begin{document}

\title{A Class of Exact  Solutions of the Faddeev Model\\}

\author{Minoru Hirayama}
 \email{hirayama@sci.toyama-u.ac.jp}
\author{Chang-Guang Shi}
 \email{shicg@jodo.sci.toyama-u.ac.jp}
\affiliation{Department of Physics, Toyama University, Gofuku 3190,
 Toyama 930-8555, {\bf Japan}\\}

\begin{abstract}
A class of exact solutions of the Faddeev model, that is, the
 modified $SO(3)$ nonlinear $\sigma$ model with the Skyrme term, is 
 obtained in the four dimensional Minkowskian spacetime. The solutions are interpreted as the isothermal coordinates of a
 Riemannian surface. One special solution of the static vortex type is investigated numerically. It is also shown that the Faddeev model is
 equivalent to the mesonic sector of the  $SU(2)$ Skyrme model where the
 baryon number current vanishes.
\end{abstract}

\maketitle

\section{\label{sec:Introduction}Introduction\protect\\}

The $SO(3)$ Faddeev model \cite{LFaddeev,FaddeevNiemi} is defined by the Lagrangian density
\begin{align}
{\cal L}_F=c_2(\partial_{\mu}\bm{n})\cdot(\partial^{\mu}\bm{n})-2c_4F_{\mu\nu}F^{\mu\nu},
\end{align}
where $\bm{n}(x)=\left(n^1(x), n^2(x), n^3(x)\right)$ is a three-component
scalar field satisfying
\begin{align}
\bm{n}^2=n^{\alpha}n^{\alpha}=1,
\label{eqn:orn}
\end{align}
$F_{\mu\nu}(x)$ is given by
\begin{align}
F_{\mu\nu}=\frac{1}{2}\bm{n}\cdot\left(\partial_{\mu}\bm{n}\times\partial_{\nu}\bm{n}\right),
\end{align}
and $c_2$ and $c_4$ are constants. It is also called the 
Skyrme-Faddeev
model or the Faddeev-Niemi model.

The field equation
\begin{align}
\partial_{\mu}(c_2\bm{n}\times\partial^{\mu}\bm{n}-2c_4F^{\mu\nu}\partial_{\nu}\bm{n})=0
\end{align}
can be expressed as 
\begin{align}
\partial_{\mu}[c_2\bm{B}^{\mu}+c_4(\bm{B}^{\mu}\times\bm{B}^{\nu})\times \bm{B}_\nu]=0,
\label{eqn:fa}
\end{align}
where $\bm{B}_{\mu}(x)$ is defined by
\begin{align}
\bm{B}_{\mu}=\bm{n}\times\partial_\mu\bm{n}.
\label{eqn:orb}
\end{align}
It is known numerically \cite{Battye} that this model possesses soliton solutions of
knot structure. They are expected to describe glueballs. For example,on
the basis of the Faddeev model, the spectrum of  glueballs is
conjectured \cite{Niemi} to be
$E_n=E_0 n^{3/4}$, 
$E_0=1500$Mev,
$n=1,2,3...$,
where $n$ is the value of the Hopf charge which classifies the mappings
from $S^3$ to $S^2$. In contrast with the rich numerical solutions, no
exact analytic solution reflecting the effects of the $c_4$-term in
${\cal{L}}_F$ is known.

On the other hand, the $SU(2)$ Skyrme model is defined by the Lagrangian
density \cite{Skyrme}
\begin{align}
{\cal L}_S=&-4c_2
 \mbox{tr}\left[(g^{\dagger}\partial_{\mu}g)(g^{\dagger}\partial
^{\mu}g)\right]+\frac{c_4}{2}\mbox{tr}\left([g^{\dagger}\partial_{\mu}g\,,\,g^{
\dagger}\partial_{\nu}g][g^{\dagger}\partial^{\mu}g\,,\,g^{\dagger}\partial^{\nu
}g]\right),
\label{eqn:Lagrangian}
\end{align}
where $g(x)$ is an element of $SU$(2).
 If we define $\bm{A}_{\mu}(x)$ by
\begin{align}
\bm{A}_{\mu}=(A_\mu^1,A_\mu^2,A_\mu^3),\quad
A_{\mu}^{\alpha}=\frac{1}{2i}\mbox{tr}\left(\tau^{\alpha}g^{\dagger}\partial_
{\mu}g\right), \label{eqn:Amug} 
\end{align}
with  $\tau_{\alpha}\hspace{1mm}(\alpha=1, 2, 3)$ being the Pauli matrices, the
field equation becomes
\begin{align}
\partial_{\mu}[c_2 \bm{A}^{\mu}+c_4(\bm{A}^{\mu}\times\bm{A}^{\nu})\times\bm{A}_{\nu}]=0.
\label{eqn:sk}
\end{align}

This equation represents the conservation of the isospin current of the
 model. Another important current of the Skyrme model is the baryon
 number current $N^\lambda(x)$ defined by
\begin{align}
N^{\lambda}=\frac{1}{12\pi^2}\varepsilon^{\lambda\mu\nu\rho}(\bm{A}_{\mu}\times\bm{A}_{\nu})\cdot\bm{A}_{\rho}.
\label{eqn:deb}
\end{align}
The conservation law
\begin{align}
\partial_{\lambda}N^\lambda=0
\end{align}
follows solely from the definition of $N^\lambda(x)$. The numerical
 analysis of the Skyrme model \cite{Sutcliffe} revealed that it possesses polyhedral
 soliton solutions with nonvanishing values of the baryon number.

The purpose of the present paper is twofold. We first clarify the
 relationship between the $SO(3)$ Faddeev model and the $SU(2)$ Skyrme
 model. We shall show that, at least at the classical level, the $SO(3)$ Faddeev model is equivalent to
 the mesonic sector of the $SU(2)$ Skyrme model where the baryon number
 current vanishes. We next explore some exact analytic solutions of the
 $SO(3)$ Faddeev model. We obtain  the solutions which are functions of
 the three variables $k\cdot x$, $l\cdot x$ and $m\cdot x$, where $k$,
 $l$ and $m$ are lightlike 4-momenta. 
 It is known \cite{Skyrme1,Doring,Belavin} that the 2-dimensional $SO(3)$ $\sigma$ model
 defined by the Lagrangian
$\mathcal{L}=(\partial_{\mu}\bm{n})\cdot(\partial^{\mu}\bm{n})$
and the
 constraint $\bm{n}^2=1$ can be solved with the help of the analytic
 functions of the complex variable $x+iy$, where $x$ and $y$ are the
 coordinates of the 2-dimensional space.

 In our solutions of the
 $SO(3)$ Faddeev model, we shall find that the isothermal coordinates of
 a Riemannian surface play an important role. They are harmonic functions on a Riemannian surface and are described in terms of an arbitrary analytic function of a complex variable. On the process of solving the field equation, we have a freedom to introduce an arbitrary function of a real variable. Then, our solutions involve an arbitrary real function, an arbitrary complex analytic function and several arbitrary parameters. For a special choice of them, we obtain a static vortex solution.

This paper is organized as follows. In Sec.\ref{sec:relationship}, we  show the equivalence
of the classical $SO(3)$ Faddeev model to the  mesonic sector of the $SU(2)$ Skyrme model
where the baryon number current vanishes  everywhere. In sec.\ref{sec:sol},
 we
 describe the Ansatz and the procedure to obtain the solutions of the $SO(3)$ Faddeev model. In Sec.\ref{sec:special cases}, some special cases excluded in the discussion of Sec.\ref{sec:sol} are discussed. 
In Sec.\ref{sec:numerical}, a special case is discussed numerically and we find that there indeed exists a nontrivial vortex configuration of $\bm{n}(x)$. 
 Sec.\ref{sec:sum} is
devoted to a summary and discussion.

\section{\label{sec:relationship}relationship between $SO(3)$ Faddeev Model and $SU(2)$
Skyrme model\protect\\} 
As is seen from Eqs. (\ref{eqn:fa}) and (\ref{eqn:sk}), the field equation
of the $SO(3)$ Faddeev model takes the same form as that of the $SU(2)$
Skyrme model. From the definition  (\ref{eqn:Amug}) for $\bm{A}_{\mu}$, we have
\begin{align}
\partial_{\mu}\bm{A}_{\nu}-\partial_{\nu}\bm{A}_{\mu}=2\bm{A}_{\mu}\times\bm{A}_{\nu},
\label{eqn:fac}
\end{align}
which should be regarded as the existency condition for $g(x)$. From the definition (\ref{eqn:orb}) of $\bm{B}_\mu$ and the condition (\ref{eqn:orn}) for
$\bm{n}$, we have
\begin{align}                                     
\partial_{\mu}\bm{B}_{\nu}-\partial_{\nu}\bm{B}_{\mu}=2\bm{B}_{\mu}\times\bm{B}_{\nu},
\label{eqn:e22}
\end{align}
which is of the same form as (\ref{eqn:fac}). Furthermore, since
$\bm{B}_{\mu}\times\bm{B}_{\nu}$ is parallel to $\bm{n}$ and
$\bm{B}_\rho$ is perpendicular to $\bm{n}$, we have
\begin{align}
\varepsilon^{\lambda\mu\nu\rho}(\bm{B}_{\mu}\times\bm{B}_{\nu})\cdot\bm{B}_{\rho}=0.
\label{eqn:ave}
\end{align}
Comparing the above constraint with the definition (\ref{eqn:deb}) of the baryon
number current for the Skyrme model, we conclude that the $SO(3)$
Faddeev model can be regarded as the mesonic sector of the $SU(2)$ Skyrme model where
the baryon number current vanishes everywhere. The degree of freedom of the Faddeev model is two, while that of the Skyrme model is three. We see that the three degrees of freedom of the Skyrme model constrained by the condition $N^\lambda(x)=0$ constitute the two degrees of freedom of the Faddeev model. We note that the above correspondence between the two models is lost for the symmetry groups bigger than $SU(2)$ and $SO(3)$. We note that the interrelation between the Faddeev and the Skrme models was discussed in Refs. \cite{Kundu, Baal} in a context different from the above.

\section{\label{sec:sol}Solving the field equation\protect\\}
In a recent paper, Yamashita and the present authors presented some classes
of solutions of the $SU(2)$ Skyrme model \cite{HY,HSY}. As for the $SU(3)$
Skyrme model, Su \cite{Su} developed the analogous method to solve the
field equation. Here we apply the method of \cite{HSY} to  
the $SO(3)$ Faddeev model. It should be noted that the method suggested
in \cite{HY} does not give a single-valued $\bm{n}(x)$.
\subsection{Ans\"atze for $\bm{B}_{\mu}$}
Just as in \cite{HSY}, we assume that $\bm{n}(x)$ depends on $x$ through the combinations $k\cdot x$, $l\cdot x$ and $m\cdot x$, where $k$, $l$ and $m$ are Minkowskian lightlike 4-momenta. Then it is convenient to express $\bm{B}_\mu(x)$ as
\begin{align}
\bm{B}_\mu=\frac{{k_\mu}\bm{a}}{\kappa^1}+\frac{{l_\mu}\bm{b}}{\kappa^2}+\frac{{m_\mu}\bm{c}}{\kappa^3},
\label{eqn:bu}
\end{align}
where 
$\kappa^i\hspace{2mm}(i=1,2,3)$ are defined by
\begin{align}
\kappa^i&=\sqrt{\frac{c_4}{c_2}\frac{(k^i\cdot k^j)(k^i\cdot
 k^k)}{(k^j\cdot k^k)}},\hspace{2mm}(ijk)=(1,2,3),\hspace{2mm} (2,3,1)
 \hspace{2mm} (3,1,2),\nonumber\\
k^1&=k,\hspace{2mm}k^2=l,\hspace{2mm} k^3=m,
\label{eqn:par}
\end{align}
and the vectors $\bm{a}$, $\bm{b}$ and $\bm{c}$ are functions of the
variables
\begin{align} 
\xi=\frac{k\cdot x}{\kappa^1},\hspace{5mm}\eta=\frac{l\cdot
 x}{\kappa^2},\hspace{5mm}\zeta=\frac{m\cdot x}{\kappa^3}. 
\end{align}
Then the condition (\ref{eqn:e22}) becomes
\begin{align}
\frac{\partial\bm{a}}{\partial\eta}-\frac{\partial\bm{b}}{\partial\xi}
&=2(\bm{b}\times\bm{a}),\label{eqn:con1}\\
\frac{\partial\bm{a}}{\partial\zeta}-\frac{\partial\bm{c}}{\partial\xi}
&=2(\bm{c}\times\bm{a}),\label{eqn:con2}\\
\frac{\partial\bm{b}}{\partial\zeta}-\frac{\partial\bm{c}}{\partial\eta}
&=2(\bm{c}\times\bm{b}).
\label{eqn:con3}
\end{align}
If we define the vectors $\bm{D}$, $\bm{E}$ and $\bm{F}$ by
\begin{align}
\bm{D}&=\bm{a}\times(\bm{b}\times\bm{a})+\bm{a}\times(\bm{c}\times\bm{a})+\bm{c}\times(\bm{b}\times\bm{a})+\bm{b}
\times(\bm{c}\times\bm{a}),\\
\bm{E}&=\bm{b}\times(\bm{a}\times\bm{b})+\bm{b}\times(\bm{c}\times\bm{b})+\bm{c}
\times(\bm{a}\times\bm{b})+\bm{a}
\times(\bm{c}\times\bm{b}), \label{eqn:DEF}\\
\bm{F}&=\bm{c}\times(\bm{a}\times\bm{c})+\bm{c}\times(\bm{b}\times\bm{c})+\bm{a}
\times(\bm{b}\times\bm{c})+\bm{b}
\times(\bm{a}\times\bm{c}),
\end{align}
the field equation (\ref{eqn:fa}) is written as
\begin{align}
&\left(\frac{\partial}{\partial\eta}+\frac{\partial}{\partial\zeta}\right)\left(
\bm{a}+\bm{D}\right)+\left(\frac{\partial}{\partial\xi}+\frac{\partial}{\partial\zeta}
\right)\left(\bm{b}+\bm{E}\right)+\left(\frac{\partial}{\partial\xi}+
\frac{\partial}{
\partial\eta}\right)\left(\bm{c}+\bm{F}\right)=0.
\label{eqn:EQMotion1}
\end{align}    
To take the condition ({\ref{eqn:ave}) into account, we assume that $\bm{a}$,
$\bm{b}$ and $\bm{c}$ are related by
\begin{align}
\bm{a}=P\bm{b}+Q\bm{c}.
\label{eqn:e10}
\end{align}
Although the condition (\ref{eqn:ave}) allows $P$ and $Q$ to depend on $x$, we hereafter assume that $P$ and $Q$ are constants.
Then the  conditions (\ref{eqn:con1}) and (\ref{eqn:con2}) become
\begin{align}
{\cal D}\bm{b}={\cal D}\bm{c}=0,
\end{align}
where ${\cal D}$ is defined by
\begin{align}
&{\cal D}=\frac{\partial}{\partial{\xi}}-P\frac{\partial}{\partial{\eta}}-Q\frac{\partial}{\partial{\zeta}}.
\end{align}
Thus, $\bm{b}$ and $\bm{c}$ are functions of
\begin{align}
\omega=(P+Q)\xi-(\eta+\zeta),\quad
\end{align}
and
\begin{align}
\omega'=Q\eta-P\zeta
\end{align}
which satisfy 
\begin{align}
{\cal D}\omega={\cal D}\omega'=0.
\end{align}
The field equation becomes
\begin{align}
&\frac{\partial\bm{X}}{\partial{\eta}}+\frac{\partial\bm{Y}}{\partial{\zeta}}=0,
\label{eqn:mes}
\\
&\bm{X}=2P\bm{b}+(P+Q+1)\bm{c}+S\bm{c}\times(\bm{b}\times\bm{c}),\\
&\bm{Y}=(P+Q+1)\bm{b}+2Q\bm{c}+S\bm{b}\times(\bm{c}\times\bm{b}),
\end{align}
where $S$ is given by
\begin{align}
S=(P+Q+1)^2-4PQ.
\end{align}
Since all the quantities considered hereafter are functions of $\omega$
and $\omega'$, the derivatives with respect to $\eta$ and $\zeta$ should
be regarded as 
\begin{align}
&\partial_\eta=\frac{\partial}{\partial\eta}=\frac{\partial \omega}{\partial
 \eta}\frac{\partial}{\partial
 \omega}+\frac{\partial \omega'}{\partial \eta}\frac{\partial}{\partial
 \omega'}=-\frac{\partial}{\partial\omega}+Q\frac{\partial}{\partial\omega'},\\
&\partial_\xi=\frac{\partial}{\partial\zeta}=\frac{\partial \omega}{\partial
 \zeta}\frac{\partial}{\partial
 \omega}+\frac{\partial \omega'}{\partial \zeta}\frac{\partial}{\partial
 \omega'}=-\frac{\partial}{\partial\omega}-P\frac{\partial}{\partial\omega'}.
\end{align}
\subsection{Reduction of the field equation}
From the definitions (\ref{eqn:orb}) and (\ref{eqn:bu}), we have
\begin{align}
\bm{b}=\bm{n}\times\partial_\eta\bm{n},\quad
 \bm{c}=\bm{n}\times\partial_\zeta
\bm{n}.
\end{align}
Eq. (\ref{eqn:con3}) is then automatically satisfied. We now express $\bm{n}(x)=\bm{n}(\omega,\omega')$ as
\begin{align}
\bm{n}=\biggl(\frac{2u}{|f|^2+1},\frac{2v}{|f|^2+1},\frac{|f|^2-1}{|f|^2+1}\biggr)
\label{eqn:e23}
\end{align}
in terms of a complex function
\begin{align}
f(\omega,\omega')=u(\omega,\omega')+iv(\omega,\omega').
\end{align} 
If we define $\kappa(\omega,\omega')$, $\lambda(\omega,\omega')$,
$\mu(\omega,\omega')$ and $\nu(\omega,\omega')$ by
\begin{align}
 &\kappa=\bm{n}\cdot(\partial_\eta\bm{n}\times\partial_\zeta\bm{n})=\frac{4}{(|f|+1)^2}\text{Im}[(\partial_\eta
 f)(\partial_\zeta \bar{f})],\label{eqn:e25}\\
&\lambda=(\partial_\eta\bm{n})^2=\frac{4}{(|f|+1)^2}(\partial_\eta
 f)(\partial_\eta \bar{f}),\label{eqn:e26}\\
&\mu=(\partial_\zeta\bm{n})^2=\frac{4}{(|f|+1)^2}(\partial_\zeta
 f)(\partial_\zeta \bar{f}),\label{eqn:e27}\\
&\nu=\partial_\eta\bm{n}\cdot\partial_\zeta\bm{n}=\frac{4}{(|f|+1)^2}\text{Re}[(\partial_\eta
 f)(\partial_\zeta \bar{f})],
\label{eqn:e28}
\end{align}
we obtain the relations
\begin{align}
&\kappa^2+\nu^2=\lambda\mu
\label{eqn:lm},\\
&\partial_\eta\bm{n}\times\partial_\zeta\bm{n}=\kappa\bm{n},\\
&\bm{b}=\frac{\lambda}{\kappa}\partial_\zeta\bm{n}-\frac{\nu}{\kappa}\partial_\eta\bm{n},\\
&\bm{c}=\frac{\nu}{\kappa}\partial_\zeta\bm{n}-\frac{\mu}{\kappa}\partial_\eta\bm{n}.
\end{align}
Then the field equation (\ref{eqn:mes}) becomes
\begin{align}
\bm{W}&\equiv\frac{\partial\bm{X}}{\partial\eta}+\frac{\partial\bm{Y}}{\partial\zeta}\nonumber\\
&=\partial_\eta(F\partial_\zeta\bm{n}-G\partial_\eta\bm{n})+\partial_\zeta(H\partial_\zeta\bm{n}-I\partial_\eta\bm{n})\nonumber\\
&=0
\end{align}
with
\begin{align}
&F=2P\frac{\lambda}{\kappa}+(P+Q+1)\frac{\nu}{\kappa}+S\kappa,\\
&G=2P\frac{\nu}{\kappa}+(P+Q+1)\frac{\mu}{\kappa},\\
&H=2Q\frac{\nu}{\kappa}+(P+Q+1)\frac{\lambda}{\kappa},\\
&I=2Q\frac{\mu}{\kappa}+(P+Q+1)\frac{\nu}{\kappa}+S\kappa.
\end{align}
Except for trivial cases, the equation $\bm{W}=0$ is equivalent to the
equations
\begin{align}
\bm{n}\cdot\bm{W}=\partial_\zeta\bm{n}\cdot\bm{W}=\partial_\eta\bm{n}\cdot\bm{W}=0.
\end{align}
It can be seen that the first equation $\bm{n}\cdot\bm{W}=0$ is
identically satisfied. With the help of the formulas
$2\partial_\eta\bm{n}\cdot\partial_\eta\partial_\zeta\bm{n}=\partial_\zeta
\lambda$,
$2\partial_\eta\bm{n}\cdot\partial_\zeta\partial_\zeta\bm{n}=2\partial_\zeta\nu-\partial_\eta\mu$,
etc., the last two equations become as
\begin{align}
&2(\partial_\eta F+\partial_\zeta H)\nu-2(\partial_\eta G+\partial_\zeta
 I)
\lambda+(F-I)\partial_\zeta\lambda
\nonumber\\
&\hspace{6mm}-G\partial_\eta\lambda-H\partial_\eta\mu+2H\partial_\zeta\nu=0,
\label{eqn:eq1}\\
&2(\partial_\eta F+\partial_\zeta H)\mu-2(\partial_\eta G+\partial_\zeta
 I)\nu+(F-I)\partial_\eta\mu\nonumber\\
&\hspace{6mm}+G\partial_\zeta\lambda+H\partial_\zeta\mu-2G\partial_\eta\nu=0.
\label{eqn:eq2}
\end{align} 
Thus we have obtained three equations (\ref{eqn:eq1}), (\ref{eqn:eq2}) and 
(\ref{eqn:lm}) for four
quantities $\kappa$, $\lambda$, $\mu$ and $\nu$. For simplicity, we
hereafter consider the case
\begin{align}
\nu=0.
\label{eqn:e41}
\end{align}
and $\kappa\le 0$. Then we have
\begin{align}
\kappa=-\sqrt{\lambda\mu}.
\end{align}
After some manipulations, we see that Eqs.(\ref{eqn:eq1}) and (\ref{eqn:eq2}) are simplified to
\begin{align}
(P+Q+1)\partial_\eta\mu+\partial_\zeta\biggl(Q\mu-P\lambda+\frac{S}{2}\lambda\mu\biggr)=0,
\label{eqn:eq3}\\
(P+Q+1)\partial_\zeta\lambda+\partial_\eta\biggl(-Q\mu+P\lambda+\frac{S}{2}\lambda\mu\biggr)=0,
\label{eqn:eq4}
\end{align}
which are the simultaneous first order differential equations for $\mu$
and $\lambda$ with constant coefficients. It is obvious that the cases
in which $P+Q+1$ and/or $S$ vanish is particularly simple.
We first discuss the case $PQ(P+Q)(Q-P-1)(Q-P+1)(3P+Q+1)(3Q+P+1)(P+Q+1)S\ne 0$
 in this section. The $S=0$ case and the $P+Q+1=0$ case will be discussed in the next section.

The solution of Eqs. (\ref{eqn:eq3}) and (\ref{eqn:eq4}) are given by
\begin{align}
&\mu=\partial_\zeta(\alpha\partial_\eta+\partial_\zeta)J,\\
&\lambda=\partial_\eta(\beta\partial_\zeta+\partial_\eta)J
\end{align}
with
\begin{align}
\alpha=\frac{2P}{P+Q+1},\quad \beta=\frac{2Q}{P+Q+1},
\end{align}
where $J$ is a function of $\omega$ and $\omega'$ 
satisfying the second order differential equation 
\begin{align}
&{\cal K}(r,s,t)\nonumber\\
&\equiv\gamma[(\alpha\partial_\eta\partial_\zeta+\partial_\zeta\partial_\zeta)J][(\beta\partial_\eta\partial_\zeta+\partial_\eta\partial_\eta)J]\nonumber\\
&\hspace{3mm}+(\alpha\partial_\eta\partial_\eta+2\partial_\eta\partial_\zeta+\beta\partial_\zeta\partial_\zeta)J\nonumber\\
&=\gamma\bigl\{(\alpha+1)r+[\alpha (P-Q)+2P]s
+(P^2-\alpha PQ)t\bigr\}\nonumber\\
&\hspace{1mm}\times\bigl\{(\beta+1)r+[\beta(P-Q)-2Q]s+(Q^2-\beta PQ)
t\bigr\}\nonumber\\
&\hspace{1mm}+\bigl[(\alpha+\beta+2)r+2(P-Q)s
+(\beta P^2-2PQ+\alpha Q^2)t\bigr]\nonumber\\
&=\text{const.}\equiv R.
\label{eqn:e48}
\end{align}
In the above equation, $\gamma$, $r$, $s$ and $t$ are defined by
\begin{align}
&\gamma=\frac{S}{P+Q+1},\\
&r=\frac{\partial^2 J}{\partial\omega^2},\quad 
s=\frac{\partial^2
J}{\partial\omega\partial\omega'},\quad
{t}=\frac{\partial^2
J}{\partial{\omega'}^2}.
\end{align}
\subsection{Intermediate integral of ${\cal K}(r,s,t)=R$}
The intermediate integral of the equation ${\cal K}(r,s,t)=R$ can be obtained 
as follows. 
Assuming that there exists an intermediate integral of the form
\begin{align}
&q-ap-b\omega-c\omega'=0,
\label{eqn:ini}\\
&p=\frac{\partial J}{\partial\omega},\quad q=\frac{\partial J}{\partial\omega'},\end{align} 
with $a$, $b$ and $c$ being constants, $r$, $s$ and $t$ are related by
\begin{align}
&r=\frac{1}{a}(s-b),\\
&t=as+c.
\end{align}
Then we have
\begin{align}
&{\cal K}\biggl(\frac{s-b}{a},s,as+c\biggr)\nonumber\\
&={\cal L}s^2+{\cal M}s+{\cal N}\nonumber\\
&=R,
\label{eqn:krd}
\end{align}
where ${\cal L}$, ${\cal M}$ and ${\cal N}$ are constant coefficients depending on the parameters $P$, $Q$ and $R$. Especially, ${\cal L}$ is independent of $R$ and is given by
\begin{align}
{\cal L}=-\cfrac{\gamma}{a^2}(aP+1)(aQ-1)\left[a(P-\alpha Q)+\alpha+1\right]\left[a(\beta P-Q)+\beta+1\right].
\end{align}
The
 constants $a$, $b$ and $c$ are now   determined so as to maintain
\begin{align}
{\cal L}={\cal M}=0,\quad {\cal N}=R.
\end{align}
It turns out that there are four sets of solutions: 
\newline
(i)
\begin{align}
a&=-\frac{1}{P},\nonumber\\
b&=\frac{\alpha(2-\alpha\beta+\gamma R)Q-(\alpha\beta+\gamma R)P}{2\gamma(\alpha\beta-1)P(P+Q)},\nonumber\\
c&=\frac{\gamma R+\alpha(2+\beta -\alpha\beta+\gamma R)}{2\gamma(\alpha\beta-1)P(P+Q)},
\end{align}
(ii)
\begin{align}
a&=\frac{1}{Q},\nonumber\\
b&=\frac{\beta(\alpha\beta-2-\gamma R)P+(\alpha\beta+\gamma R)Q}{2\gamma(\alpha\beta-1)Q(P+Q)},\nonumber\\
c&=\frac{\gamma R+\beta(2+\alpha-\alpha\beta+\gamma R)}{2\gamma(\alpha\beta-1)Q(P+Q)},
\end{align}
(iii)
\begin{align}
&a=\frac{\alpha+1}{\alpha Q-P}=\frac{3P+Q+1}{P(Q-P-1)},\nonumber\\
&b=\frac{2\alpha Q-\alpha\beta P+\gamma P R}{2\gamma(P-\alpha Q)(P+Q)},\nonumber\\
&c=\frac{\alpha(\beta+2)-\gamma R}{2\gamma(P-\alpha Q)(P+Q)},
\end{align}
(iv)
\begin{align}
&a=\frac{\beta+1}{Q-\beta P}=\frac{3Q+P+1}{Q(Q-P+1)},\nonumber\\
&b=\frac{2\beta P-\alpha\beta Q+\gamma Q R}{2\gamma(\beta P-Q)(P+Q)},\nonumber\\
&c=\frac{-\beta(\alpha+2)+\gamma R}{2\gamma(\beta P-Q)(P+Q)}.
\end{align}
\subsection{Solution of the intermediate integral}

We have seen that there exist rather simple intermediate integrals of Eq. (\ref{eqn:e48}). We next obtain their general solutions, that is, the solutions of Eq. (\ref{eqn:ini}) containing an arbitrary function.  If we set $J(\omega,\omega')$ as
\begin{align}
&J=-\frac{b}{2a}\omega^2+\frac{c}{2}{\omega'}^2+\psi(z),
\label{eqn:e62}
\\
&z=\frac{\omega}{a}+\omega'
\label{eqn:psi}
\end{align}
with $\psi(z)$ being an  arbitrary function of $z$, we easily see that 
Eq. (\ref{eqn:ini}) is satisfied. Thus we have obtained a class of solutions of Eq. (\ref{eqn:e48}). For the above $J(\omega,\omega')$, $\lambda$ and $\mu$ become as 
\begin{align}
&\lambda=A\psi''(z)+B,\\
&\mu=C\psi''(z)+D,
\end{align}
where $A$, $B$, $C$ and $D$ are constants given by
\begin{align}
&A=\frac{\alpha+1}{a^2}+\frac{\alpha(P-Q)+2P}{a}+P(P-\alpha Q),\\
&B=-\frac{(\alpha+1)b}{a}+P(P-\alpha Q)c,\\
&C=\frac{\beta+1}{a^2}+\frac{\beta(P-Q)-2Q}{a}+Q(Q-\beta P),\\
&D=-\frac{(\beta+1)b}{a}+Q(Q-\beta P)c.
\end{align}
We note that $A$ and $C$ are given by
\newline
(i) $A=0, \hspace{3mm} C=(P+Q)^2$,
\newline
(ii) $A=(P+Q)^2,\hspace{3mm}C=0$,
\newline
(iii) $A=0,\hspace{2mm} C=\frac{(P+Q)^2 S}{(3P+Q+1)^2}$,
\newline
(iv) $A=\frac{(P+Q)^2 S}{(3Q+P+1)^2},\hspace{1mm}C=0,$
\newline
for the allowed four values $a=-\frac{1}{P}$, $\frac{1}{Q}$, $\frac{3P+Q+1}{P(Q-P-1)}$, $\frac{3Q+P+1}{Q(Q-P+1)}$, respectively.
\subsection{Geometric meaning of $u$ and $v$}
From the assumption
\begin{align}
\nu=\frac{4}{(u^2+v^2+1)^2}\left[(\partial_\eta u)(\partial_\zeta u)
+(\partial_\eta
 v)(\partial_\zeta
 v)\right]=0,
\label{eqn:e71}
\end{align}
we have
\begin{align}
\frac{\partial_\eta u}{\partial_\zeta
 v}=-\frac{\partial_\eta v}{\partial_\zeta
 u}\equiv\rho.
\end{align}
Then the definitions (\ref{eqn:e25}), (\ref{eqn:e26}) and (\ref{eqn:e27}) lead us to the relation
\begin{align}
\rho=-\frac{\lambda}{\kappa}=-\frac{\kappa}{\mu}=\sqrt{\frac{\lambda}{\mu}}.
\end{align}
For the functions $\lambda$ and $\mu$ obtained in the previous subsection, we see that $\rho$ is a function of $z$ and is determined by $\psi''(z)$. It is remarkable that $u$ and $v$ satisfy
\begin{align}
(\partial_\eta+i\rho\partial_\zeta)(u+iv)=0,
\label{eqn:eq5}
\end{align}
or more explicitly
\begin{align}
\biggl[\biggl(\frac{\partial}{\partial\omega}-Q\frac{\partial}{\partial\omega'}\biggr)+i\rho\biggl(\frac{\omega}{a}+\omega'\biggr)\biggl(\frac{\partial}{\partial\omega}+P\frac{\partial}{\partial\omega'}\biggr)\biggr]\left[u(\omega,\omega')+i v(\omega,\omega')\right]=0.
\label{eqn:e75}
\end{align}
To see the geometric meaning of $u$ and $v$, we consider the Riemannian surface ${\cal S}$ whose first fundamental form is given by
\begin{align} 
ds^2=g_{\alpha\beta}dx^\alpha dx^\beta=\rho
d\eta^2+\frac{1}{\rho}d\zeta^2
\end{align}
with
\begin{align}
&dx^1=d\eta=-\frac{Pd\omega-d\omega'}{P+Q},\\
&dx^2=d\zeta=-\frac{Qd\omega+d\omega'}{P+Q},\\
&g_{11}=\rho,\hspace{2mm}g_{22}=\frac{1}{\rho},\hspace{2mm}g_{12}=g_{21}=0.
\end{align}
Defining $g^{\alpha\beta}$ by
\begin{align}
g^{11}=\frac{1}{\rho},\hspace{2mm}g^{22}={\rho},\hspace{2mm}g^{12}=g^{21}=0,
\end{align}
the first Beltrami operator $\Delta_1$ and the
 second Beltrami operator $\Delta_2$ on ${\cal S}$ are given by
\begin{align}  
&\Delta_1(j)= g^{\alpha\beta}\frac{\partial j}{\partial
x^\alpha}\frac{\partial j}{\partial
 x^\beta}=\frac{1}{\rho}(\partial_\eta j)^2+\rho(\partial_\zeta j)^2,\\
&\Delta_1(j,k)= g^{\alpha\beta}\frac{\partial j}{\partial
x^\alpha}\frac{\partial k}{\partial
 x^\beta}=\frac{1}{\rho}(\partial_\eta j)(\partial_\eta
 k)+\rho(\partial_\zeta j)(\partial_\zeta k),\\
&\Delta_2(j)=\frac{1}{\sqrt g}\frac{\partial}{\partial
 x^\beta}\biggl[
{\sqrt{g}} g^{\alpha\beta}\frac{\partial j}{\partial
 x^\alpha}\biggr]
=
\partial_\eta\biggl(\frac{\partial_\eta j}{\rho}\biggr)
+\partial_\zeta
(\rho\partial_\zeta j).
\end{align}

It is now straightforward to obtain
\begin{align}
&\Delta_1(u)=\Delta_1(v),\quad
\Delta_1(u,v)=0,\\
&\Delta_2(u)=\Delta_2(v)=0.
\end{align}
Thus $u$ and $v$ are harmonic functions on the surface ${\cal S}$. In terms of the variables $u$ and $v$, $ds^2$ is 
expressed as
\begin{align}
ds^2=K(du^2+dv^2),\hspace{5mm}
K=\frac{1}{\Delta_1(u).} 
\end{align}
The variables $u$ and $v$ with the above property are called the isothermal
coordinates of ${\cal S}$. As was mentioned in Sec.\ref{sec:Introduction}, the two-dimensional $SO(3)$ nonlinear
$\sigma$ model 
 is solved in terms of the variables $\tilde{u}(x,y)$ and $\tilde{v}(x,y)$ satisfying
the Cauchy-Riemann relation 
\begin{align}
\biggl(\frac{\partial}{\partial
x}+i\frac{\partial}{\partial y}\biggr)(\tilde{u}+i\tilde{v})=0.
\end{align}
In our case, however, the solutions of the field equation with higher
order terms such as $\bm{c}\times(\bm{b}\times\bm{c})$ in $\bm{X}$ and
$\bm{b}\times(\bm{c}\times\bm{b})$ in $\bm{Y}$ are expressed by the
isothermal coordinates of the surface ${\cal S}$. We note that the surface ${\cal S}$
is fixed when the second fundamental form as well as the first one are specified.

\subsection{Solutions of Eq. (\ref{eqn:eq5})} 
To obtain $u(\omega,\omega')$ and $v(\omega,\omega')$, we must solve
Eq. (\ref{eqn:eq5}) with a given $\rho(z)$ which is a function of $z$. The solution
is obtained through two steps. As the first step, we
observe that $u_1(\omega,\omega')$ and $v_1(\omega,\omega')$ given by
\begin{align}
&u_1=u_0(Q\omega+{\omega'})+\sigma(z),\label{eqn:e88}\\
&v_1=v_0(-P\omega+{\omega'})+\chi(z),\label{eqn:e89}
\end{align}
with $u_0$ and $v_0$ being constants satisfy the equation
\begin{align}
\biggl[\biggl(\frac{\partial}{\partial\omega}-Q\frac{\partial}{\partial\omega'}\biggr)+i\rho(z)\biggl(\frac{\partial}{\partial\omega}+P\frac{\partial}{\partial\omega'}\biggr)\biggr]\left(u_1+i
 v_1\right)=0
\end{align}
if $\sigma$ and $\chi$ satisfy
\begin{align}
\biggl(\frac{1}{a}-Q\biggr)\sigma'(z)=\biggl(\frac{1}{a}+P\biggr)\rho(z)\chi'(z)
\end{align}
and
\begin{align}
({P}+{Q})[u_0\rho(z)-v_0]+\biggl(\frac{1}{a}+P\biggr)\rho(z)\sigma'(z)+\biggl(\frac{1}{a}-Q\biggr)\chi'(z)=0.
\end{align}
Recalling that there are four allowed values of $a$, we obtain
\newline
Case(i):
\begin{align}
&a=-\frac{1}{P}\\
&\sigma(z)=\text{const.},\\
&\chi(z)={u_0}\int^{z}\rho(z')dz'-v_0 z.
\end{align}
Case(ii):
\begin{align}
&a=\frac{1}{Q}\\
&\sigma(z)={v_0}\int^{z}\frac{1}{\rho(z')}dz'-u_0z,\\
&\chi(z)=\text{const},
\end{align}
Case (iii) or (iv):
\begin{align}
&a=\frac{3P+Q+1}{P(Q-P-1)}\hspace{1mm}\text{or}\hspace{1mm} \frac{3Q+P+1}{Q(Q-P+1)},\\
&\sigma(z)=-{a(P+Q)(1+aP)}
\int^z
 \frac{\rho(z')(
u_0\rho(z')-v_0)}{(1+aP)^2\rho(z')^2+(1-aQ)^2}dz',\\
&\chi(z)=-{a(P+Q)(1-aQ)}\int^z
 \frac{
u_0\rho(z')-v_0}{(1+aP)^2\rho(z')^2+(1-aQ)^2}dz'.
\end{align}
As the second step, we set $u(\omega,\omega')$ and $v(\omega,\omega')$
as
\begin{align}
&u(\omega,\omega')=\text{Re}f(u_1+iv_1)\equiv U(u_1,v_1),\\
&v(\omega,\omega')=\text{Im}f(u_1+iv_1)\equiv V(u_1,v_1),
\end{align}
where $F(u_1+iv_1)$ is an analytic function of the complex variable
$u_1+iv_1$ and hence $U(u_1,v_1)$ and $V(u_1,v_1)$ satisfy the
Cauchy-Riemann relation
\begin{align}
\frac{\partial U}{\partial u_1}=\frac{\partial V}{\partial v_1},\quad
 \frac{\partial U}{\partial v_1}=-\frac{\partial V}{\partial u_1}.
\end{align}
Then we have
\begin{align}
&\biggl[\biggl(\frac{\partial}{\partial\omega}-Q\frac{\partial}{\partial\omega'}
\biggr)+i\rho(z)\biggl(\frac{\partial}
{\partial\omega}+P\frac{\partial}{\partial\omega'}\biggr)\biggr]\left(u+i
 v\right)\nonumber\\
&=\biggl(\frac{\partial U}{\partial u_1}+i\frac{\partial V}{\partial
 u_1}\biggr)
\biggl[\biggl(\frac{\partial}{\partial\omega}-Q\frac{\partial}{\partial\omega'}
\biggr)+i\rho(z)\biggl(\frac{\partial}
{\partial\omega}+P\frac{\partial}{\partial\omega'}\biggr)\biggr]\left(u_1+i
 v_1\right)\nonumber\\
&=0.
\end{align}
Thus we have obtained $u(\omega,\omega')$ and  $v(\omega,\omega')$ which
involve the parameters $u_0$ and $v_0$ and an arbitrary analytic function
$F$. Then $\bm{n}$ is constructed through Eq. (\ref{eqn:e23}). The case in which $(P+Q)(Q-P-1)(Q-P+1)(3P+Q+1)(3Q+P+1)$ vanishes can be discussed analogously. 
\section{\label{sec:special cases}special cases\protect\\}
Up to now, we have been assuming that both $S$ and $P+Q+1$ are
nonvanishing. In this section, we discuss briefly the cases that $S$ or 
 $P+Q+1$ vanishes.  
\subsection{The $S=0$ case}
In this case, we have $\alpha\beta-1=\gamma=0$ and Eqs. (\ref{eqn:eq4}) and 
(\ref{eqn:eq3}) become
\begin{align}
&\alpha^2\biggl(\frac{\partial}{\partial\omega}+P\frac{\partial}{\partial\omega'}
\biggr)
 \lambda=\biggl[(2\alpha+1)\frac{\partial}{\partial\omega}+(P-2\alpha Q)
\frac{\partial}
{\partial\omega'}\biggr]\mu,\\
&\biggl(\frac{\partial}{\partial\omega}-Q\frac{\partial}{\partial\omega'}\biggr)
\mu=\biggl[(\alpha^2+2\alpha)\frac{\partial}{\partial\omega}+(2\alpha P-\alpha^2Q)
\frac{\partial}
{\partial\omega'}\biggr]\lambda.
\end{align}
The solution of this system of equations are given by
\begin{align}
&\lambda=G(z_1)+\omega H(z_1),\\
&\mu=-\frac{P}{Q}\lambda-\frac{3P+Q+1}{Q(P+Q+1)(P-Q+1)}\int^{z_1}G(z)dz,\\
&z_1=\frac{\omega}{a_1}+{\omega'}\label{eqn:lam1},\\
&a_1=-\frac{3P+Q+1}{P(P-Q+1)}=\frac{2(P-Q)}{P+Q+1},\\
\label{eqn:mu1}
\end{align}
where $G(z_1)$ and  $H(z_1)$ are arbitrary functions of $z_1$. If
$H(z_1)$ is set equal to zero, $\rho=\sqrt{{\lambda}/{\mu}}$ becomes
a function of the variable $z_1$. Then, we can obtain $u$ and $v$ and
hence $\bm{n}$ through the method in  In sec. \ref{sec:sol}. 
\subsection{The $P+Q+1=0$ case}
In this case, we have $S=-4PQ=4P(P+1)$ and Eqs. (\ref{eqn:eq4}) and 
(\ref{eqn:eq3}) become
\begin{align}
&\biggl(\frac{\partial}{\partial\omega}-Q\frac{\partial}{\partial\omega'}\biggr)\left[(P+1)\mu+P\lambda+2P(P+1)\lambda\mu\right]=0,\\
&\biggl(\frac{\partial}{\partial\omega}+P\frac{\partial}{\partial\omega'}\biggr)\left[(P+1)\mu+P\lambda-2P(P+1)\lambda\mu\right]=0.
\end{align}
The solution of these equations is given by
\begin{align}
&(P+1)\mu+P\lambda=\frac{1}{2}\left[J(Q\omega+\omega')+K(-P\omega+\omega')\right],\\
&P(P+1)\lambda\mu=\frac{1}{4}\left[J(Q\omega+\omega')-K(-P\omega+\omega')\right],
\end{align}
where $J(Q\omega+\omega')$ and $K(-P\omega+\omega')$ are arbitrary
functions of the variables $Q\omega+\omega'$ and $-P\omega+\omega'$,
respectively. If we set $J(Q\omega+\omega')$ or $K(-P\omega+\omega')$
equal to zero, $\rho=\sqrt{\lambda/\mu}$ is a function of the variable
$-P\omega+\omega'$ or $Q\omega+\omega'$. Then, we can obtain $u$ and $v$
by the method of  sec. \ref{sec:sol}.
\newpage
\section{\label{sec:numerical}numerical investigation of a special case\protect\\}
To understand the behaviour of $\bm{n}$, we here consider the following
specific case:
\begin{align}
&k=(k_0,k_0,0,0),\hspace{2mm}l=(l_0,0,l_0,0),\hspace{2mm}l=(m_0,0,0,m_0),\nonumber\\
&P=Q=1,\hspace{2mm}R=0,\nonumber\\
&u_0=1,\hspace{2mm}v_0=-1.
\end{align}
Then we have
\begin{align}
&\kappa^1=\sqrt{\frac{c_4}{c_2}}\hspace{1mm}k_0,\hspace{2mm}\kappa^2=\sqrt{\frac{c_4}{c_2}}\hspace{1mm}l_0,\hspace{2mm}\kappa^3=\sqrt{\frac{c_4}{c_2}}\hspace{1mm}m_0,\nonumber\\
&\xi=\sqrt{\frac{c_2}{c_4}}(x_0-x_1),\hspace{2mm}\eta=\sqrt{\frac{c_2}{c_4}}(x_0-x_2),\hspace{2mm}\zeta=\sqrt{\frac{c_2}{c_4}}(x_0-x_3).
\end{align}
In this case, the variables $\omega$ and $\omega'$ are independent of
the time variable $x_0$:
\begin{align}
&\omega=\sqrt{\frac{c_2}{c_4}}(-2x_1+x_2+x_3),\label{eqn:5.3}\\
&\omega'=\sqrt{\frac{c_2}{c_4}}(-x_2+x_3).\label{eqn:5.4}
\end{align}
We also specify the functions $\psi(z)$ and $F(u_1+iv_1)$ as
\begin{align} 
&\psi(z)=z^2,\\
&F(u_1+iv_1)=u_1+iv_1.
\end{align}
Then we have
\begin{align}
&\rho=\frac{1}{\sqrt{5}}\\
&u=u_1=\frac{1}{49}[(51+2\sqrt{5})\omega+(39-10\sqrt{5})\omega'],\\
&v=v_1=\frac{1}{49}[(34-3\sqrt{5})\omega+(26+15\sqrt{5})\omega'].
\end{align}
and
\begin{align}
|f|^2=u^2+v^2=\frac{26}{49}(3\omega^2+4\omega\omega'+3\omega'^2).
\end{align}
The static energy density 
\begin{align}
E=4c_2\frac{\nabla f\cdot\nabla f^{*}}{(1+|f|^2)^2}-4c_4\frac{(\nabla
 f\times\nabla f^{*})^2}{(1+|f|^2)^4}
\end{align}
turns out to be 
\begin{align}
&E=\frac{c_2^2}{c_4}\varepsilon,\nonumber\\
&\varepsilon=\frac{4(\delta_1+\delta_2)}{(1+|f|^2)^2}+\frac{16(\delta_1\delta_2-\delta_3^2)}{(1+|f|^2)^4},
\end{align}
where $\delta_1$, $\delta_2$ and $\delta_3$ are given by
\begin{align}
&\delta_1=6\biggl(\frac{\partial u_1}{\partial\omega}\biggr)^2+2\biggl(\frac{\partial
 u_1}{\partial\omega'}\biggr)^2=\frac{8(353-6\sqrt{5})}{343},\\
&\delta_2=6\biggl(\frac{\partial v_1}{\partial\omega}\biggr)^2+2\biggl(\frac{\partial
 v_1}{\partial\omega'}\biggr)^2=\frac{8(193+6\sqrt{5})}{343},\\
&\delta_3=6\frac{\partial u_1}{\partial\omega}\frac{\partial v_1}{\partial\omega}+2\frac{\partial
 u_1}{\partial\omega'}\frac{\partial v_1}{\partial\omega'}=\frac{4(384+5\sqrt{5})}{343},
\end{align}
We see that $\bm{n}$ is equal to $(0,0,-1)$ at $(\omega,\omega')=(0,0)$ and
that $\bm{n}$ approaches to $(0,0,1)$ as $|\omega|$ and $|\omega'|$
increase. In Figs.1, 2, 3, 4 and 5, the behaviour of $\bm{n}$ is
shown. In Fig.6, the direction of $\bm{n}$ is shown schematically. In
Fig.7, the behaviour of $\varepsilon$ is shown. As is seen from Eqs. (\ref{eqn:5.3}) and (\ref{eqn:5.4}), the ($\omega$, $\omega'$)-plane can be
regarded as a plane with the normal $(1,1,1)/\sqrt{3}$ in the
physical $(x_1,x_2,x_3)$-plane.

\section{\label{sec:sum}Summary and
discussion\protect\\}

In this paper, we have clarified the interrelation between the Skyrme
model for the matrix field $g(x)\in SU(2)$ and the Faddeev model for the
isovector scalar field $\bm{n}(x)\in S^2$. By comparing the vector field
descriptions making use of
$\bm{A}_\mu=\frac{1}{2i}\text{tr}(\bm{\tau}g^{\dagger}\partial_\mu g)$ and
$\bm{B}_\mu=\bm{n}\times\partial_\mu \bm{n}$, it was concluded that the
Faddeev model can be regarded as the mesonic sector of the Skyrme model
in which the baryon number current vanishes everywhere.

Next, we have explored the exact solutions of the Faddeev model. Under
the Ansatz that $\bm{n}(x)$ is a function of the variables $k\cdot x$,
$l\cdot x$ and $m\cdot x$ with $k$, $l$ and $m$ being Minkowskian
lightlike 4-vectors, the field equation for $\bm{B}_\mu$ has been
reduced to the nonlinear differential equation (\ref{eqn:EQMotion1}) for the isovector
scalar fields $\bm{a}$, $\bm{b}$ and $\bm{c}$. From the assumption 
(\ref{eqn:e10})
which ensures the condition (\ref{eqn:ave}), We have seen that $\bm{a}$,$\bm{b}$ and
$\bm{c}$ should depend only on the two variables $\omega$ and
$\omega'$. We have seen that the field equation for $\bm{n}$ can be
rewritten as the coupled equations (\ref{eqn:eq1}), (\ref{eqn:eq2}) and (\ref{eqn:lm}) for the scalars $\kappa$,
$\lambda$, $\mu$ and $\nu$ defined by Eqs. (\ref{eqn:e25}), (\ref{eqn:e26}), (\ref{eqn:e27}) and
(\ref{eqn:e28}). Restricting ourselves to the case of vanishing $\nu$, the above equations have been
reduced to the tractable system of equations (\ref{eqn:eq3}) and (\ref{eqn:eq4}) for $\lambda$ and
$\mu$. We have found that these equations can be reduced to the
nonlinear equation (\ref{eqn:e48}) for $J(\omega,\omega')$, which has the solution (\ref{eqn:e62}) involving one arbitrary function $\psi(z)$. The restriction
(\ref{eqn:e71}) has led us to the generalized Cauchy-Riemann relation (\ref{eqn:eq5}), which has
been solved in terms of an arbitrary analytic function $F(u_1+iv_1)$
where $u_1$ and $v_1$ are given by (\ref{eqn:e88}) and (\ref{eqn:e89}). Thus, roughly speaking,
the field equation of the Faddeev model has been solved under the
assumptions (\ref{eqn:bu}), (\ref{eqn:e10}) and (\ref{eqn:e41}). We have pointed out that $u$ and $v$ in the
$\nu=0$ case can be interpreted as a pair of isothermal coordinates of a
Riemannian surface.
Our solutions of the Faddeev model involve arbitrary lightlike 4-momenta $k$, $l$ and $m$, arbitrary parameters $P$, $Q$,
$R$, $u_0$ and $v_0$, an arbitrary function $\psi(z)$, and an arbitrary
analytic function $F(u_1+iv_1)$. 

As an example of a numerical
estimation, we have considered the simplest case $P=1$, $Q=1$, $R=0$, $u_0=-v_0=1$,
$\psi(z)=z^2$ and $F(u_1+iv_1)=u_1+iv_1$ and found the static vortex  solution in which
$\bm{n}=(0,0,-1)$ at $(\omega,\omega')=(0,0)$ and
$\bm{n}=(0,0,1)$ for large $\omega^2+{\omega'}^2$. 
If we define the topological charge $\Phi$ by
\begin{align}
\Phi=\frac{1}{4\pi}\int_{-\infty}^{\infty}d\omega\int_{-\infty}^{\infty}d\omega'
\bm{n}\cdot\biggl(\frac{\partial\bm{n}}{\partial\omega}\times\frac{\partial\bm{n}}
{\partial\omega'}\biggr),
\end{align}
$\Phi$ is given by
\begin{align}
\Phi=-mn,
\end{align}
where $m$ and $n$ are the winding numbers of the two mappings
$(\omega,\omega')\rightarrow (u_1,v_1)$
and $(u_1,v_1)\rightarrow(u,v)$, respectively. The former mapping is
governed by the arbitrary function $\rho(z)$, while the latter by the
arbitrary analytic function $F(u_1+iv_1)$.

We hope our method gives some hints to obtain the analytic expressions
of the knot solutions in \cite{Battye} found by numerical investigations 
. Since $\bm{n}$ appears in the description of Yang-Mills
field \cite{Cho, Morita, Faddeevn}, the vortex structure of $\bm{n}$
observed here might suggest the same structure of the Yang-Mills
field. We should note that the existence of the vortex solution of the Faddeev model was also discussed by Kundu and Rybakov \cite{KunduRy}. We finally note that, although no exact analytic expression was
presented, the vortex solution of the Abelian Higgs model was found by
Nielsen and Olesen \cite{Nielsen} thirty years ago.

\begin{acknowledgments}
The authors are grateful to Prof. A. Kundu for correspondence and to their colleagues for kind interests in this work. 
\end{acknowledgments}


\newpage

\begin{figure}[htb]
\includegraphics[width=10cm]{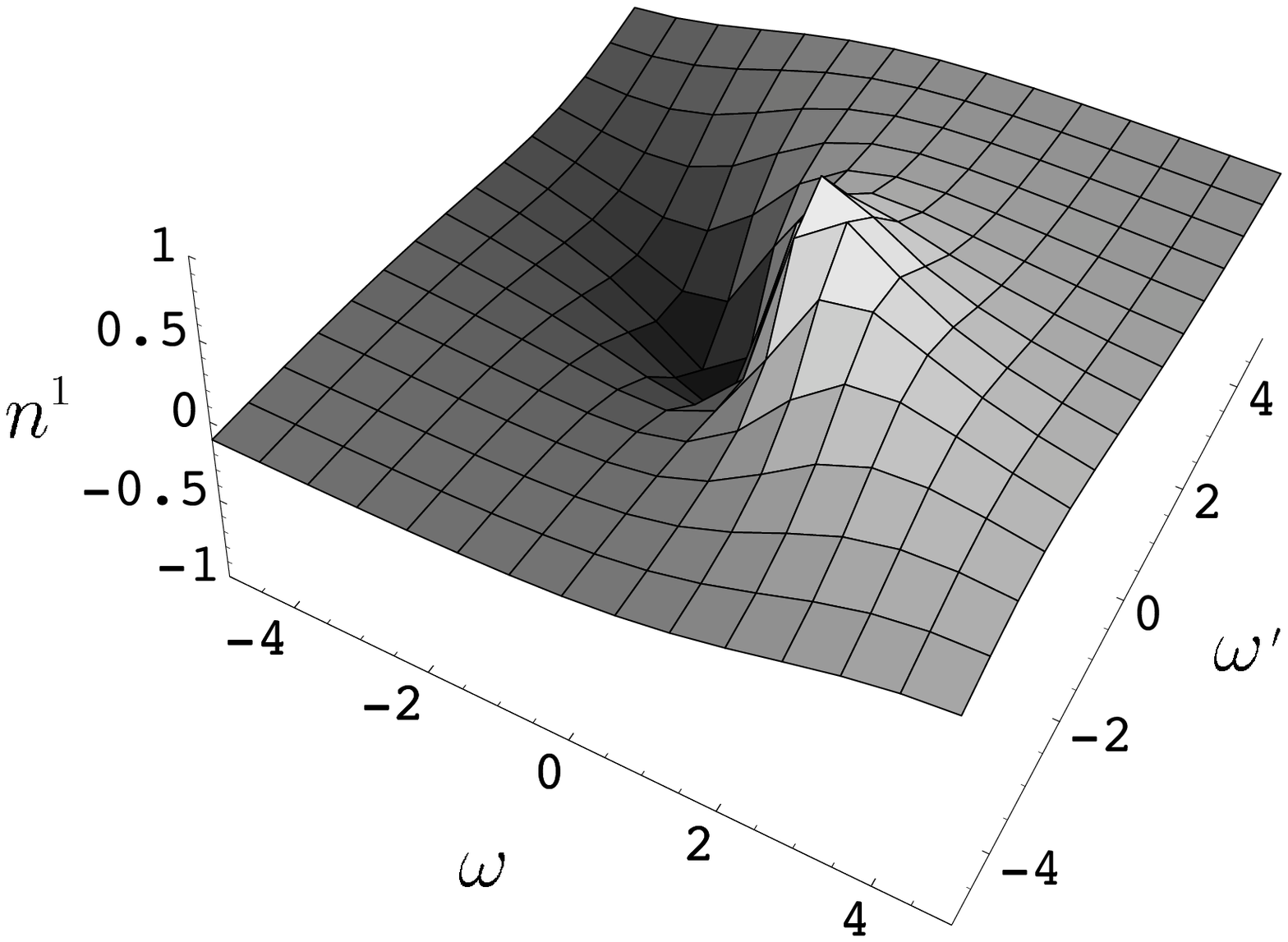}
\caption{The behaviour of $n^1(\omega,\omega')$.}
\end{figure}
 \begin{figure}[htb]
\includegraphics[width=12cm]{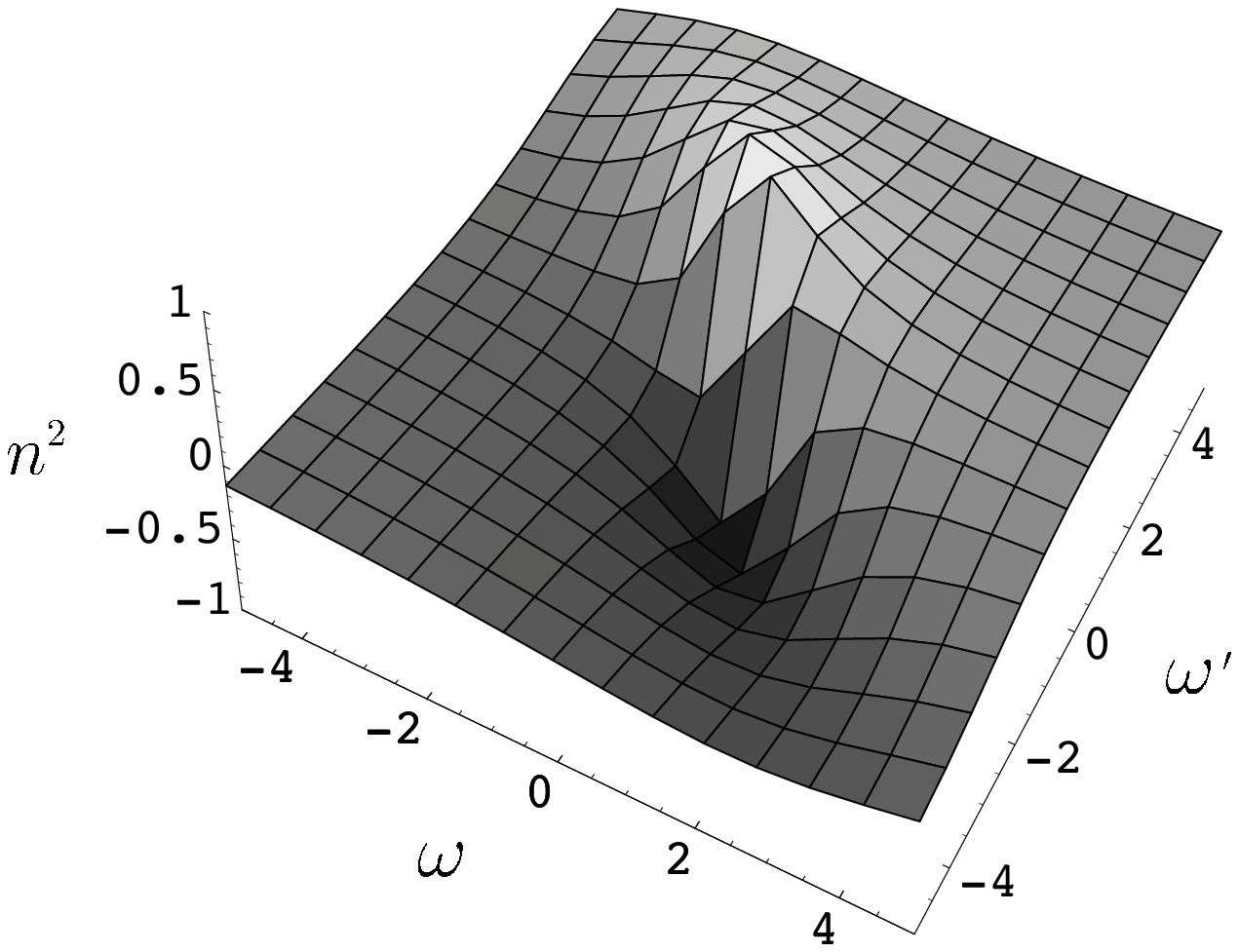}
\caption{The behaviour of $n^2(\omega,\omega')$.}
\end{figure}
\begin{figure}[htb]
\includegraphics[width=10cm]{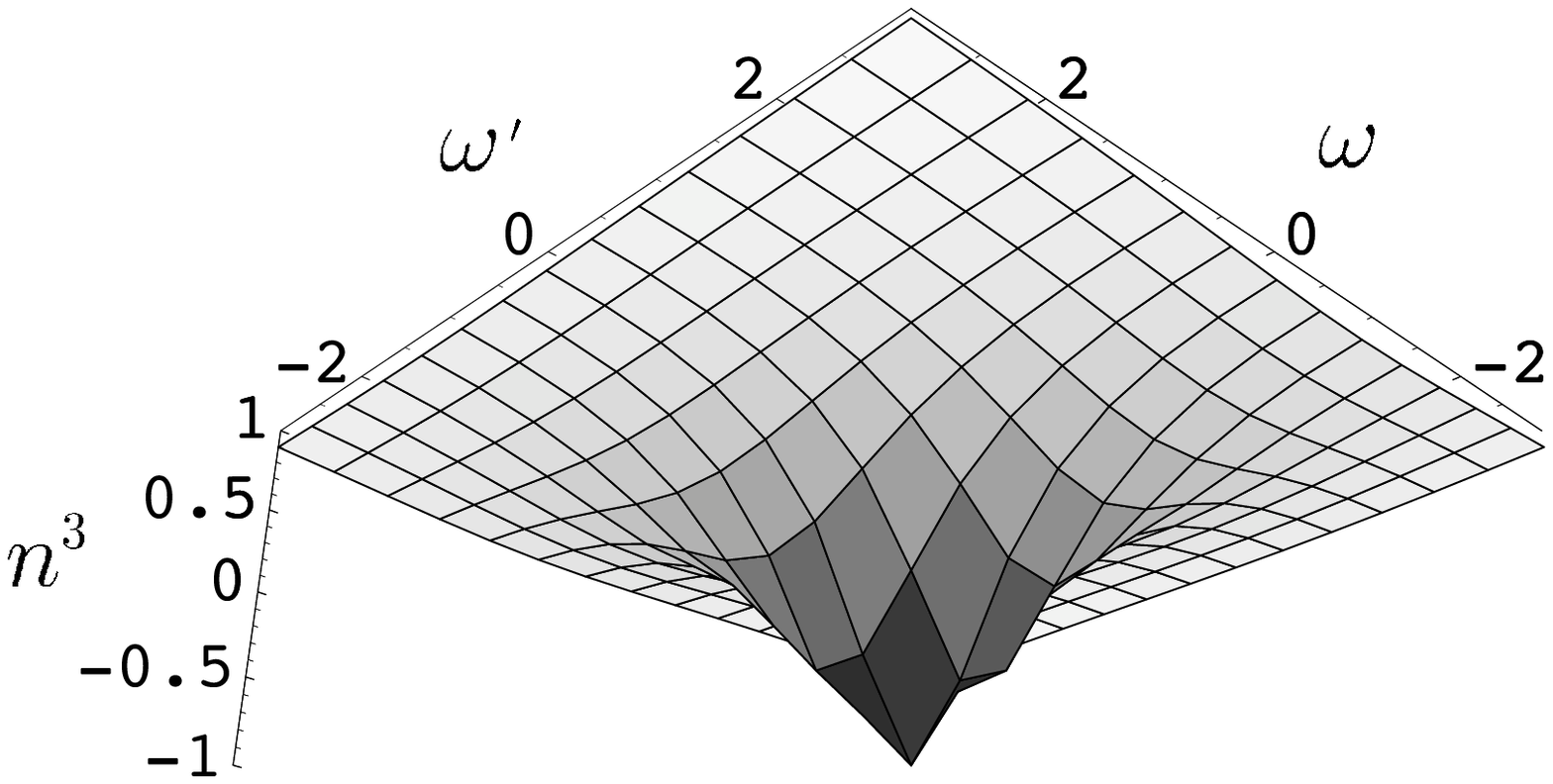}
\caption{The behaviour of $n^3(\omega,\omega')$ looked at from the bottom.}
\end{figure}
\begin{figure}[htb]
\includegraphics[width=10cm]{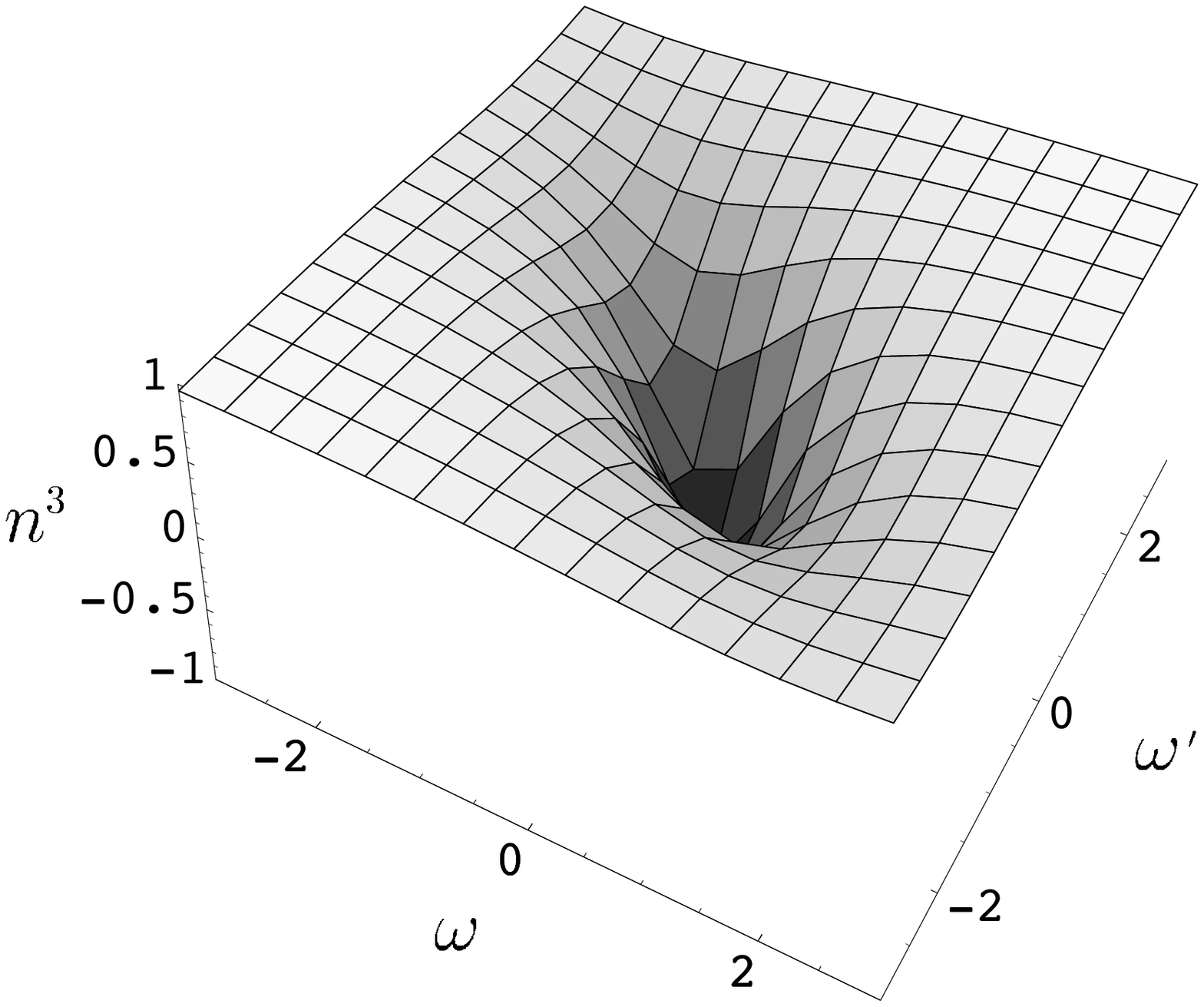}
\caption{The behaviour of $n^3(\omega,\omega')$ looked at from the top.}
\end{figure}
\begin{figure}[htb]
\includegraphics[width=12cm]{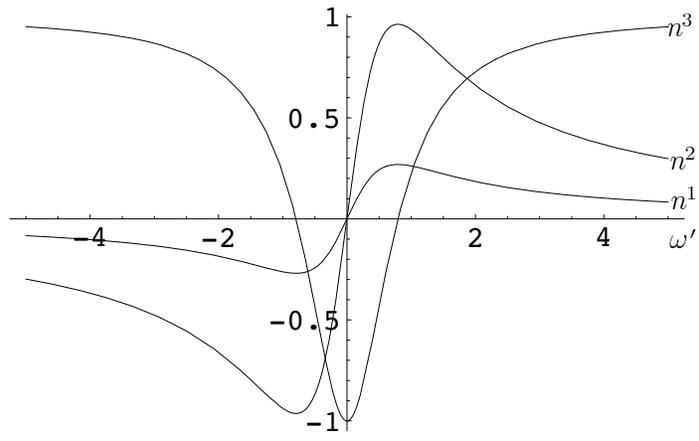}
\caption{The behaviours of $n^1(\omega,\omega')$,
$n^2(\omega,\omega')$ and $n^3(\omega,\omega')$ for $\omega=0$.}
\end{figure}
\begin{figure}[htb]
\includegraphics[width=10cm]{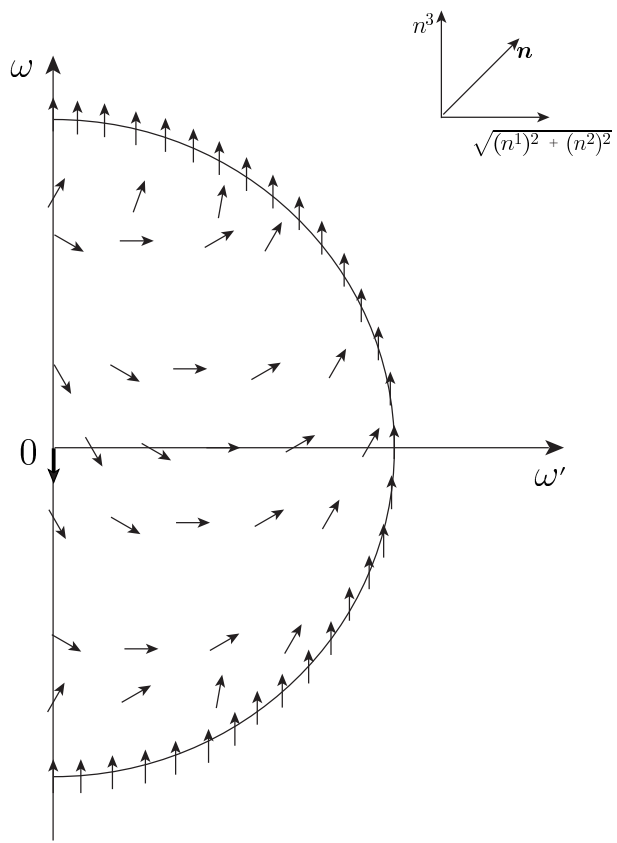}
\caption{The schematic description of the direction of
$\bm{n}(\omega,\omega')$.}
\end{figure}
\begin{figure}[htb]
\includegraphics[width=12cm]{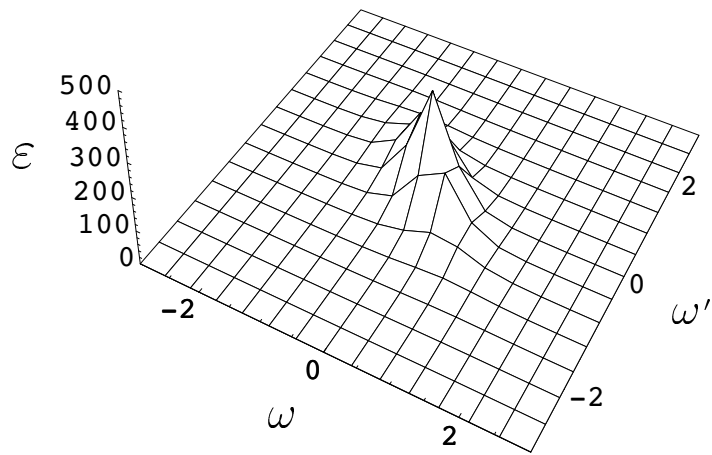}
\caption{The behaviour of $\varepsilon(\omega,\omega')$.}
\end{figure}

\end{document}